
\input harvmac
\def\journal#1&#2(#3){\unskip, \sl #1~\bf #2 \rm (19#3) }
\def\nextline{\unskip\nobreak\hskip\parfillskip\break}
     
\def\brst{Q_{brst}}
\def\zbar{{\bar z}}
\def\ZZ{{\bf Z}}
\Title{\vbox{\baselineskip12pt{\hbox{TIT/HEP-172(revised)}
\hbox{September 1991}}}}
{\vbox{
\centerline{On Moduli Space of C=0 Topological Conformal Field Theories}
}}
\vfill
\centerline{Hisahiro Yoshii $^{1,2}$}
\bigskip{\baselineskip14pt
\centerline{Department of Physics, Tokyo Institute of Technology,}
\centerline{Meguro-ku, Tokyo 152, JAPAN}}
\bigskip
\bigskip
\bigskip
\noindent
We studied the marginal deformation of the $c=0$ topological
conformal field theories (TCFT).
We showed that
topological $SL(2)$ Wess-Zumino-Witten (WZW) model,
topological superconformal ghost system,
TCFT constructed from the $N=2$ superconformal system and
two dimensional topological gravity
belong to the same one parameter family
(moduli space) of the $c=0$ TCFT's.
We conjectured that the $N=2$ TCFT constructed from
the Wolf space realization of $N=4$ superconformal algebra
belongs to another family.

\vfill
\centerline{(To appear in {\sl Physics Letters}{\bf B})}

\footnote{}{$^1$ JSPS fellow.}
\footnote{}{$^2$ E-mail address: hyoshii@cc.titech.ac.jp,
yoshii@jpnrifp}

\Date{September 16, 1991}
%
%
\subsec{Introduction}

Topological natures of a quantum field theory are sometimes wove
into the supersymmetric ground state.
If a system has unbroken $N=2$ supersymmetry one obtains the
non-trivial supercharge cohomology
\ref\Witten{E. Witten\journal Nucl. Phys. &B202 (82) 253-316.}.
In such a system, one can construct a topological field theory
by interpreting it as the BRST cohomology.
Topological field theories become much more important
through the study of the two and three dimensional quantum gravities.
In two dimension the system with $N\geq2$ superconformal symmetry
has the chiral ring structure
\ref\LVW{W. Lerche, C. Vafa and N.P. Warner\journal Nucl. Phys. &B324 (89)
427-474.}
which contains the topological informations.
The chiral states are in fact regarded as the BRST invariant states
of the TCFT constructed from the original theory
\ref\EY{T. Eguchi and S. K. Yang\journal Mod. Phys. Lett. A &5 (90)1693-1701.}
\ref\yoshii{H. Yoshii\journal Phys. Lett. B &259 (91) 279-284.},
in which one of the supercurrent plays a role of the BRST current
due to the Felder's work
\ref\Felder{G. Felder\journal Nucl. Phys. &B317 (89) 215--236.}.

Here we discuss the marginal deformation of the TCFT.
The marginal deformation is one of the recipes to
investigate the phase structure.
In two dimensional conformally invariant system,
marginal operators $m_i(z, \zbar)$
are primary fields with conformal weight ($1, 1$).
Consider the marginal perturbation by adding the following term in the action
\eqn\action{\delta S=\sum_i{\delta g_i\over2\pi}\int d^2z m_i(z, \zbar)\,.}
In path-integral formalism the addition of the term \action\ corresponds
to the insertion of $m_i(z,\zbar)$ into the correlation functions.
If primary fields $\psi_\alpha$ with conformal weights
($h_\alpha,\bar h_\alpha$) satisfy the following algebra
\eqn\algebra{m_i(z, \zbar)\psi_\alpha(w, \bar w)
\sim \sum_\beta c_{i\alpha\beta}(z-w)^{h_\beta-h_\alpha-1}
(\zbar-\bar w)^{\bar h_\beta-\bar h_\alpha-1}\psi_\beta(w, \bar w)\,,}
the perturbation shifts the conformal weights
($h_\alpha,\bar h_\alpha$) of the operator $\psi_\alpha$ by
\eqn\weights{\delta h_\alpha=\delta\bar h_\alpha=-\sum_i c_{i\alpha\alpha}
\delta g_i\,.}

The marginal operator $m_i$ that changes no conformal structure is said to
be integrable, which satisfies $c_{iij}=0$ for any marginal operators $m_j$.
If such an operator exists, perturbation by this operator gives
the one parameter family of theories.
This space is interpreted as the moduli space of classical solutions to
the string equations of motion.
The $c=1$ and $c=3/2$ cases have been extensively studied in
\ref\GDVV{
P. Ginsparg\journal Nucl. Phys. B &295 [FS21] (88) 153-170.\nextline
R. Dijkgraaf, E. Verlinde and H. Verlinde
\journal Commun. Math. Phys. &115 (88) 649-690.\nextline
L. Dixon, P. Ginsparg and J. Harvey\journal Nucl. Phys.  &B306 (88)470-496.}.
The $c=3k/k+2$ ($k\in\ZZ$) case is partially studied in
\ref\Yang{S. K. Yang\journal Phys. Lett. B&209 (88) 242-246.},
where the $SU(2)$ WZW theory and the minimal $N=2$ superconformal
theory are shown to belong to the same moduli space.

%
%
\subsec{Topological first-order system.}

We discuss first the supersymmetric first-order system
\ref\FMS{D. Friedan, E. Martinec and S. Shenker
\journal Nucl. Phys. &B271 (86) 93-165.}
\eqn\action{S={1\over\pi}\int d^2z(b\bar\del c+\beta\bar\del \gamma),}
where ($\beta, \gamma$) and ($b, c$) are conjugate fields with
bosonic and fermionic statistics, respectively.
Their operator products are given by
\eqn\ghost{b(z)c(w)={1\over z-w},\quad \beta(z)\gamma(w)={-1\over z-w}.}
To specify the conformal structure of this system we assume
the stress tensor of the following form
\eqn\super{
L=-nb\del c+(1-n)(\del b)c-(n-\ha)\beta\del\gamma
+({3\over2}-n)(\del\beta)\gamma\,,
}
where $n$ is a real number.
Let us define the conformal operators
\eqn\two{
\eqalign{
G^+&=2b\gamma\,,\cr
G^-&=-(2n-2)(\del\beta)c-(2n-1)\beta\del c\,,\cr
J&=-(2n-2)bc-(2n-1)\beta\gamma\,,\cr
}}
with dimensions ${3\over2},{3\over2}$ and $1$, respectively.
Then the set of operators $(L, G^+, G^-, J)$ satisfy
the $N=2$ superconformal algebra with central charge
\eqn\center{c=9-12n\,.}
This system is a generalization of the superconformal ghost system which
is reproduced at $n=2$ and the minimal $N=2$ SC theory
($n=\ha(k+3/k+2)$, $k\in {\bf Z}_+$).

Following ref \EY\ one can construct a topological field theory from
the $N=2$ superconformal system \super.
At first we construct a new stress tensor by twisting
\super\ by the $U(1)$ current
\eqn\general{
L_\lambda=L+\ha\del J=-\lambda b\del c+(1-\lambda)(\del b)c
-\lambda\beta\del\gamma+(1-\lambda)\del\beta\gamma\,,
}
where $\lambda=2n-1$.  Then $L_\lambda$ satisfies the Virasoro algebra
with vanishing central charge.
The charge screening operator with respect to \general\ is given by
\eqn\nilpotent{\brst=\oint d\xi G^+(\xi)\,,}
which is nilpotent and is regarded as the BRST operator
in the Felder's interpretation.
Kernel of $\brst$ is nothing but the
{\it chiral} states for the $N=2$ superconformal theory\LVW.
For any value of $\lambda$, the stress tensor $L_\lambda$
is given by the BRST transformation of the operator $G^-$.
We will call this system \general\ as
the topological first-order system(TFOS).
By twisting \general,
the operators, $b$ and $\beta$ ($c$ and $\gamma$) acquire
the same conformal weights $\lambda$ ($1-\lambda$).
Since ($\beta, \gamma$) and ($b, c$) are still conjugate,
the TFOS enjoys the duality
\eqn\dual{\eqalign{
\lambda&\longleftrightarrow1-\lambda\,,\cr
(b,c)&\longleftrightarrow(c,b)\,,\cr
(\beta, \gamma)&\longleftrightarrow(-\gamma, \beta)\,.\cr}}
The self-dual point is $\lambda=\ha$.

The system \general\ has also an $SL(2)$ structure,
whose generators are given by
\eqn\KM{\eqalign{J^+&=-\gamma\,,\cr
J^3&=-\beta\gamma-b c\,,\cr
J^-&=\beta^2\gamma+2\beta b c\,. \cr}}
The $SL(2)$ Kac-Moody algebra satisfied by \KM\ has vanishing level.
These generators are also given by the BRST transformation
\eqn\exact{\eqalign{
J^+&=\ha\{\brst, -c\},\cr
J^3&=\ha\{\brst, -c\beta\},\cr
J^-&=\ha\{\brst, c\beta^2\},.\cr}}
With respect to \general,
the operators $J^+, J^3$ and $J^-$ are primary fields with
conformal weights $1-\lambda, 1$ and $1+\lambda$, respectively.
The Sugawara form constructed from \KM\ has the following form
\eqn\Virasoro{L_{SL(2)}=(\del\beta)\gamma+(\del b)c\,.}
This is just the $\lambda=0$ case of \general.
One may say that the topological $SL(2)$ Wess-Zumino-Witten model
is realized at $\lambda=0$ (and its conjugate point, $\lambda=1$).

The existence of the $SL(2)$ currents suggests the existence of the
integrable marginal operators.
At $\lambda=0$ we have five independent integrable marginal operators;
\eqn\marginal{J^3(z)J^3(\zbar)\,,\quad
\sum c_{\alpha\beta}J^\alpha(z)J^\beta(\zbar)\,.}
where $\alpha, \beta=+, -$.
In this sense $\lambda=0$ will be a multicritical
point of the theory space of this model.
Here we will mainly discuss the marginal deformation generated by $J^3$.
It is convenient to separate $J^3$ into the bosonic
and fermionic parts
\eqn\dummy{j_+=bc\,,\qquad j_-=\beta\gamma\,.}
As is well known that each current, $j_\pm$, has current anomaly
(background charge) $Q_\pm=\pm(2\lambda-1)$ with opposite sign.
To cancel them we consider the following marginal operator
\eqn\marginal{m_3(z, \zbar)=j_+(z)\bar j_+(\zbar)+j_-(z)\bar j_-(\zbar)\,,}
which satisfies the integrability condition.
The operator products of $m_3(z, \zbar)$ with primary fields are given by
\eqn\change{
m_3(z,\zbar)\pmatrix{b\cr\beta\cr}(w)
={1\over z-w}\pmatrix{b\cr\beta\cr}(w),\quad
m_3(z,\zbar)\pmatrix{c\cr\gamma\cr}(w)
={-1\over z-w}\pmatrix{c\cr\gamma\cr}(w)\,,}
which say that the marginal deformation by $m_3(z,\zbar)$
decreases the conformal weights of $b$ and $\beta$
and increases the ones of $c$ and $\gamma$ by the
same magnitude[see eq.\weights].
This change corresponds to the following deformation of the stress tensor
\eqn\deform{L_{SL(2)}\longrightarrow
L_\lambda=L_{SL(2)}+\lambda\del J^3\,.}
Hence we verified that the general supersymmetric first order system
 \general\ is obtained by the marginal perturbation of the topological
$SL(2)$ WZW model.
Since the background charge of $J^3$ vanishes, the above deformation
will correspond to the insertion of puncture operators.

Using the Coulomb gas representation (bosonization),
the above observation will be more satisfactorily understood.
Following \FMS
we define the two bosonic fields $\varphi_\pm$ by
\eqn\boson{
j_\pm(z)=\mp\del \varphi_\pm(z), \quad
<\varphi_\pm(z)\varphi_\pm(w)>=\pm\ln |z-w|\,.}
With these fields the marginal operator \marginal\ is written as follows
\eqn\coulomb{m_3(z, \zbar)=
\del_z\varphi_+\overline{\del_z\varphi_+} +
\del_z\varphi_-\overline{\del_z\varphi_-}\,.}
One may assume that the two fields $\phi_\pm=\varphi_\pm(z)+
\overline{\varphi_\pm(z)}$ are compactified on the torus.
In the case the marginal operator \coulomb\ changes the compactification
radius of $\phi_\pm$.
We must show that it also causes a change of the value of $\lambda$.
We give the stress tensor for the bosonic and fermionic system
\eqn\stress{L_\pm=\pm(\ha j_\pm^2-\ha Q_\pm\del j_\pm)\,,}
where the stress tensor $L_\pm$ has central charge
$c_\pm=1\mp3Q_\pm^2$.
[For the bosonic system, the bosonization rule \boson\ is incomplete and
one must introduce the residual $c=-2$ stress tensor commuting with
$j_-$(see \FMS). ]
The bosonized action which reproduces \stress\ reads as follows
\eqn\action{
S_\pm={1\over4\pi}\int d^2z[
\mp2\del_z\phi_\pm\del_\zbar\phi_\pm -\ha Q_\pm\sqrt{g}R\phi_\pm]\,, }
where $\sqrt{g}R$ is the Einstein lagrangian in two dimension.
The equation of motion is given by
\eqn\motion{\del_z\del_\zbar\phi_\pm=\pm{1\over8}Q_\pm\sqrt{g}R\,.}
Since $Q_\pm=\pm(2\lambda-1)$, the marginal perturbation by
$m_3(z, \zbar)$ changes the value of $\lambda$.

At the self-dual point $\lambda=\ha$,
both of the background charges $Q_\pm$ vanish.
The operators $j_+\bar j_+$ and $j_-\bar j_-$ may be regarded as
the two independent integrable marginal operators.
This fact makes the marginal perturbations at $\lambda=\ha$
be completely different from the one at the other points.
At $\lambda=\ha$ r.h.s. of eq. \motion\ vanishes and the perturbation
only changes the compactification radius of $\phi_\pm$.

Since the $SU(2)$ WZW model and the $N=2$ unitary minimal superconformal
model belong to the same moduli space of CFTs\Yang,
one might expect the same situation also occurs in the present case.
Although one will never find the topological $N=2$ SCFT in the
theory space of the TFOS.
\footnote{$^\dagger$}
{At the self-dual point of the moduli space, $\lambda=\ha$,
the untwisted $N=2$ superconformal theory \super\
has vanishing central charge.
Although the theory cannot be written as the BRST exact form
and is not regarded as the topological theory.}
A topological $N=2$ CFT is obtained by twisting the $N=4$ SCFT,
whereas the maximal symmetry of the untwisted theory \super\
is the $N=2$ superconformal symmetry.
The TFOS does not show even the $N=1$ superconformal symmetry due to that
the bosonic and the fermionic fields have the same conformal weight.

It is important to note that the $\lambda=2$ theory is
the two dimensional topological quantum gravity coupled with the $c=-2$
conformal matter
\ref\LPW{J. M. F. Labastida, M. Pernici, and E. Witten\journal
Nucl. Phys. &B310 (88) 611-624.}
\ref\Distler{J. Distler\journal Nucl. Phys. &B342 (90) 523-538.}.

%
%
\subsec{Topological N=2 Superconformal Theory.}
The topological field theory with $N=2$ superconformal symmetry
\ref\nojiri{S. Nojiri, KEK preprint 90-179 and 90-194.}
is obtained by the topological twist of the $N=4$ superconformal theory
with $so(4)$ Kac-Moody algebra
\ref\GPTP{M. G\"unaydin, J. L. Petersen, A. Taormina and A. Van Proeyen
\journal Nucl. Phys.  &B322 (89)402-430.}.
The algebra consists of a dimension 2 stress tensor $L$,
four dimension $3\over2$ supercurrents $G_a$ $(a=\pm,\pm K)$,
$su(2)\oplus su(2)\oplus u(1)$ currents
$(B^{+i}, B^{-i}, U)$ $(i=\pm, 3)$
and four free fermions $Q_a$ $(a=\pm,\pm K)$.
The central charge is written in terms of the two levels of the
$su(2)\oplus su(2)$ Kac-Moody algebra as follows
\eqn\center{c=6{k^+k^-\over k},\qquad k=k^++k^-\,.}
Denoting $k^-/k=\gamma$ the algebra is often referred as $\cal A_\gamma$.
Regarding $G^+$ as the BRST current,
the massless state condition for ${\cal A}_\gamma$ and
the BRST invariant state condition are identified.
The generators for the topological $N=2$ superconformal algebra are
defined as follows
\eqn\new{\eqalign{
&\hat L_\gamma=L+{k^-\over k}\del B^{+3}+{k^+\over k}\del B^{-3}
		+\ha (1-2\gamma)\del \hat J_\gamma\,,\cr
&\hat G^\pm_\gamma=G_{\pm K}\mp 2{k^\mp\over k}\del Q_{\pm K}\,,\cr
&\hat J_\gamma=B^{+3}-B^{-3}+U\,.\cr
}}
%
For any $0<\gamma<1$ the central charge of the algebra vanishes and
the generators $\hat L_\gamma, \hat G^\pm_\gamma$ and $\hat J_\gamma$ are
given by the BRST transformation of $G_-, B^{\pm -}$ and $Q_-$, respectively.
Using the left- and right-moving $U(1)$ currents we can construct the
integrable marginal operator $m_{N=2}=\hat J_\gamma(z)\hat J_\gamma(\bar z)$.

Here we note that any $N=4$ algebra ${\cal A}_\gamma$ is realized by the
supersymmetric GKO construction on the Wolf space
\ref\Van{
A. Van Proeyen\journal Class. Quantum Grav. &6 (89) 1501-1508.\nextline
Ph. Spindel, A. Sevrin, W. Troost and A. Van Proeyen\journal
Nucl. Phys. &B308 (88) 662-698\journal ibid. &B311 (88/89) 465-492.}.
In
\ref\EYH{T. Eguchi, S. K. Yang, S. Hosono, Tokyo Univ. preprint UT-577.}
TCFT's are constructed by twisting the general coset model.
Due to the existence of three complex structures on Wolf spaces,
their hidden $N=1$ fermionic symmetry will be enhanced to the explicit
$N=2$ superconformal symmetry and hidden $N=4$ symmetry, by the complex
structure transformation of the $N=1$ supercurrent.
If one considers the Wolf space of the form $W=G/(H\otimes SU(2)\otimes
U(1))$, the levels $k^\pm$ of the two $SU(2)$ Kac-Moody algebras
are given by
\eqn\level{k^+=n+1\,,\quad k^-=\tilde g -1\,.}
where $\tilde g$ is the dual Coxeter number of $G$ and $n$ is the
level of the $G$ Kac-Moody algebra.
The $N=4$ SCA has the same central charge to the one obtained in \EYH.

Now we consider the marginal deformation of the $N=2$ topological
conformal field theory generated by $m_{N=2}$.
Using the $N=2$ SCA one finds that the perturbation by $m_{N=2}$
decreases (increases) the conformal weight of $\hat G^+_\gamma$
($\hat G^-_\gamma$).
Since the $U(1)$ current $\hat J_\gamma$ is nilpotent
and has vanishing screening charge
the perturbation by $m_{N=2}$ will
be expressed by the following deformation of the stress tensor
\eqn\zero{\hat L_\kappa=\hat L_\gamma+\kappa\del \hat J_\gamma\,,}
where $\kappa$ is an arbitrary parameter.
Though it breaks the $N=2$ superconformal symmetry,
topological natures are still preserved.
The family of these theories enjoys the duality:
\eqn\last{\eqalign{\kappa&\longleftrightarrow-\kappa\,,\cr
\hat G^+&\longleftrightarrow\hat G^-,\cr}}
where the self-dual point $\kappa=0$ is the topological $N=2$ CFT.
At $\kappa=\pm\ha$, the theory coincides with the one obtained by twisting
the topological $N=2$ conformal theory with the second BRST current
$\hat G^\pm$.

We note that the theory contains another candidate of the
integrable marginal operator,
$\tilde m={\cal J}(z)\overline{\cal J}(\zbar)$ where
${\cal J}={k^+-k^-\over k}(B^+_3+B^-_3)+U$.
This operator is not BRST invariant and changes the conformal
weight of the BRST current $G_+$.
Perturbation by $\tilde m$ breaks the topological nature of the theory.

\bigskip
%
%
\subsec{Discussions.}
The topological conformal field theory has only primary fields with vanishing
conformal weights.
Such theory has been considered by Moore and Seiberg
\ref\MS{G. Moore and N. Seiberg\journal Commun. Math. Phys.
&123 (89)177-254.}.
In fact the topological conformal field theories belong to the special
subclass of the classical CFT in which not only conformal weights but
also all quantum numbers vanish.
Classical CFTs are group theories and
topological CFTs include only topological informations.
If their observation is applicable to the TCFTs,
the twist of the stress tensor corresponds to the classical limit and
the quantum reconstruction will be also possible.

\bigskip
\bigskip
\centerline{\bf Acknowledgments}
I would like to thank N. Sakai and S. Nojiri for discussions and comments.
Research supported in part by Grant-in-Aid
for Scientific Research for the Ministry of Education, Science and
Culture No. 02952028.

\bigskip
\bigskip
\listrefs

\bye